\documentclass[sigconf]{acmart}

\usepackage[inline]{enumitem}
\usepackage{amsmath}
\usepackage{algorithm}
\usepackage{algorithmic}
\usepackage{hyperref}
\usepackage{textcomp} 
\usepackage{bm}
\usepackage{caption}
\usepackage{subcaption}

\newtheorem{definition}{Definition}

\def\BibTeX{{\rm B\kern-.05em{\sc i\kern-.025em b}\kern-.08emT\kern-.1667em\lower.7ex\hbox{E}\kern-.125emX}}
    
\setcopyright{none}
\acmConference[DSHealth '19: 2019 KDD Workshop on Applied Data Science for Healthcare]{DSHealth '19: 2019 KDD Workshop on Applied Data Science for Healthcare}{August 4 -- 8, 2019}{Anchorage, Alaska}

\acmPrice{}
\acmDOI{}
\acmISBN{}
\copyrightyear{}

\begin{document}

%

\title{Leveraging Linguistic Characteristics for Bipolar Disorder Recognition with Gender Differences}

%

\author{Yen-Hao Huang}
\affiliation{%
  \institution{National Tsing Hua University}
  \city{Hsinchu}
  \country{Taiwan}
}
\email{yenhao0218@gmail.com}

\author{Yi-Hsin Chen}
\affiliation{%
  \institution{National Tsing Hua University}
  \city{Hsinchu}
  \country{Taiwan}
}
\email{eunicebes@gmail.com}

\author{ Fernando H. Calderon}
\affiliation{
  \institution{National Tsing Hua University}
  \city{Hsinchu}
  \country{Taiwan}
  }
\email{fhcalderon87@gmail.com}

\author{Ssu-Rui Lee}
\affiliation{%
  \institution{National Tsing Hua University}
  \city{Hsinchu}
  \country{Taiwan}
}
\email{gn01697933@gmail.com}

\author{Shu-I Wu}
\authornote{The Doctor of Medicine in Department of Medicine, Mackay Medical College.}
\affiliation{
    \institution{Mackay Memorial Hospital}
    \city{Taipei}
    \country{Taiwan}
    }
\email{shuiwu@g.ntu.edu.tw}

\author{Yuwen Lai}
\affiliation{%
  \institution{National  Chiao Tung University}
  \city{Hsinchu}
  \country{Taiwan}
  }
\email{yuwen.lai@gmail.com}

\author{Yi-Shin Chen}
\authornote{Corresponding author.}
\affiliation{%
  \institution{National Tsing Hua University}
  \city{Hsinchu}
  \country{Taiwan}
  }
\email{yishin@gmail.com}

%
\renewcommand{\shortauthors}{Huang and Chen, et al.}

%
\begin{abstract}
Most previous studies on automatic recognition model for bipolar disorder (BD) were based on both social media and linguistic features. 
The present study investigates the possibility of adopting only language-based features, namely the syntax and morpheme collocation.
We also examine the effect of gender on the results considering gender has long been recognized as an important modulating factor for mental disorders, yet it received little attention in previous linguistic models.
The present study collects Twitter posts 3 months prior to the self-disclosure by 349 BD users (231 female, 118 male).
We construct a set of syntactic patterns in terms of the word usage based on graph pattern construction and pattern attention mechanism. 
The factors examined are gender differences, syntactic patterns, and bipolar recognition performance. 
The performance indicates our F1 scores reach over 91\% and outperform several baselines, including those using TF-IDF, LIWC and pre-trained language models (ELMO and BERT). 
The contributions of the present study are: 
(1) The features are contextualized, domain-agnostic, and purely linguistic. 
(2) The performance of BD recognition is improved by gender-enriched linguistic pattern features, which are constructed with gender differences in language usage. 
\end{abstract}

%

%
\maketitle

\section{Introduction}
Bipolar disorder (BD) is a mental illness that can be characterized by recurrent episodes of mania and depression~\citep{american2013diagnostic}. It is known to be under-diagnosed and under-treated worldwide~\cite{prince2007no}. 
The assessment for BD is crucially for the treatment~\cite{sit2004women}, however,  it involves frequent visits to a health practitioner, which is restricted to the limited mental health resources. 


In the past two decades, social media have become popular platforms for self-disclosure.
Given that the voluntary self-disclosures contain rich information regarding BD patients mental and emotional states~\citep{joinson2001self}. 
Many researchers have been leveraging social media data for mental disorder studies, including candidate seeking, behavior analyses and mental illness recognition.

The existed recognition approaches could be placed in two groups:
\begin{enumerate*}
\item \textbf{Social-behavior-based model}~\cite{coppersmith2014quantifying, chang2016subconscious} was designed to quantify the social media engagements (e.g., posts per day, mention, retweets ,etc.), mood polarity, and emotion features that could also be derived from users' posts.
However, the models suffer from two major drawbacks: they can only be applied to social media data and the information of time for each post has to be incorporated which casts a more demanding processing load.
\item \textbf{Post-content-based model} extracts features from texts as well as image content of social media posts. 
To directly compare to our goals, the review below only focuses on studies on text contents. 
Two major text-based research methods are:
\begin{enumerate*}
    \item \textit{Dictionary} approaches, such as the \textbf{LIWC} dictionary~\cite{pennebaker2001linguistic} which categorizes words in different classes based on psychological usages, were used to construct recognition models for mental illnesses~\cite{coppersmith2014quantifying,  coppersmith2015adhd, sekulic2018not}. However, such a pre-defined dictionary requires expensive manual effort for maintenance and needs constant renewal; and
    \item \textit{Character language models (CLMs)}, the data-driven keywords/phrases methods, were also proposed for recognition tasks. Many studies~\cite{coppersmith2014quantifying, chang2016subconscious, huang2017detection} based on CLMs employed term frequency inverse document frequence (TF-IDF), language and topic model approaches to build the classification models. 
    These models learn the content related words/phrases in a dynamic manner to distinguish whether the given text were originated from a certain patient. 
\end{enumerate*}
However, by the content dependent nature of CLMs, it may mis-distinguish a domain expert as a patient~\cite{chang2016subconscious}, due to frequent word profile~\footnote{For example, the psychiatrist of BD may be judged as patient by frequently mention terms such as ``bipolar'' or ``medicine''.}.
\end{enumerate*}

Previous epidemiological research showed that the mood disorder rate among women is significantly higher than male~\cite{seedat2009gender}.
Researchers \cite{preoctiuc2015role, munmun2017gender} also found that there are considerable gender differences on social media disclosure for people with mental disorder.
As a result, the current research places focus on understanding the linguistic expressions of BD and investigate the possible differences between gender.
Regarding gender factor, while some social-behavior-based models combine gender information as demographic features in their models, existing pure CLMs models usually don't consider this factor. 

The present study focuses on language-based feature of the social media posts by BD users. 
We aim to establish a set of gender-specific syntactic patterns in which the data are relatively easy to be retrieved and allow for the transfer as prior knowledge to health practitioners for clinical routine. 
Our syntactic pattern representations incorporate the data-driven concept by CLMs, yet avoid the specific keyword problems for mental disorder recognition. 

\subsubsection*{\textbf{Ethics}}
The proposed model will be employed as assistive tools, if and only if, healthcare practitioners and patients agree to use it, rather than offer a diagnosis. The personal information of users in this study are masked and de-identified as the protection of privacy.

\section{Gender-Enriched Dataset}
\label{sec:dataset}
Self-disclosure allows various approaches to discover users who may have a mental illness on social media \cite{chang2016subconscious, sekulic2018not}. 
To offer gender analyses, there are two major  data collecting steps in our study which are
\begin{enumerate*}
    \item BD and control data collection; and
    \item Gender information retrieval.
\end{enumerate*}
Firstly, for BD group, tweets from users experiencing BD episodes were collected from $2012$ to $2018$, by the time-specific data collection approach~~\footnote{Users are qualified and collected if time specific information is provided in their disclosure tweets. For example, ``I was diagnosed bipolar on June, 2017''.}~\cite{huang2017detection}.  The control group was retrieved from a Kaggle dataset~\footnote{\url{https://www.kaggle.com/crowdflower/twitter-user-gender-classification}}, which included their account name and gender. Self-disclosure users were filtered out in the control group to avoid the possibility that they had a mental health condition.

\noindent
\textbf{Gender Information Retrieval}: 
As Twitter does not require users to provide the gender information, there is only a small proportion of users have demographic information. 
To avoid the biases, instead of adopting existing classifiers, the gender of a user in the BD group was labeled by manually effort based on  
\begin{enumerate*}
    \item \textbf{account name} (screen name as well), 
    \item \textbf{profile}, and
    \item \textbf{photos}.
\end{enumerate*}
After cross-examining all the rules, users are labeled with gender tags if valid. Users with contradictory gender information are discarded in this work. 

\noindent
\textbf{Onset Episode}: 
For the BD group, we only considered the $3$ months period before the diagnosis time, based on the assumption that this 2 to 3 months period contains the most important information to perform more accurate diagnoses~\cite{huang2017detection}. Similar $3$-months period were also randomly sampled for control groups.

The statistics of our dataset are summarized in Table~\ref{table:dataset}. In total, there are $231$ and $118$ female/male BD users and the median amount for tweets are $356$ and $339$ within the $3$-months period, respectively. 

\begin{table}[ht]
    \small
    \centering
    \begin{tabular}{c | c c | c c } 
        \hline
        & \multicolumn{2}{c|}{\textbf{Bipolar}} & \multicolumn{2}{c}{\textbf{Control}}\\
           & \small{\textbf{Female}} & \small{\textbf{Male}}& \small{\textbf{Female}} & \small{\textbf{Male}} \\
         \hline
         \small{\textbf{users}} & $231$ & $118$ & $1.6k$ & $2k$  \\
         \textbf{ $m$\small{ of Total}} & $6.9k$ & $5.7k$& $2.03k$ & $2.26k$  \\
         \textbf{ $\mu$ \small{of Total} } & $16.7k$ & $15.3k$& $2.07k$ & $2.21k$  \\
         \textbf{$m$ \small{of $3$-months}} & $356$ & $339$ & $454$ & $501$  \\
         \textbf{$\mu$ \small{of $3$-months}} & $1.1k$ & $719$& $806$ & $991$  \\
         \hline
    \end{tabular}
    \caption{Statistics of users and tweets.}
    \label{table:dataset}
\end{table}

\begin{figure}
    \centering
    \includegraphics[width=1\linewidth]{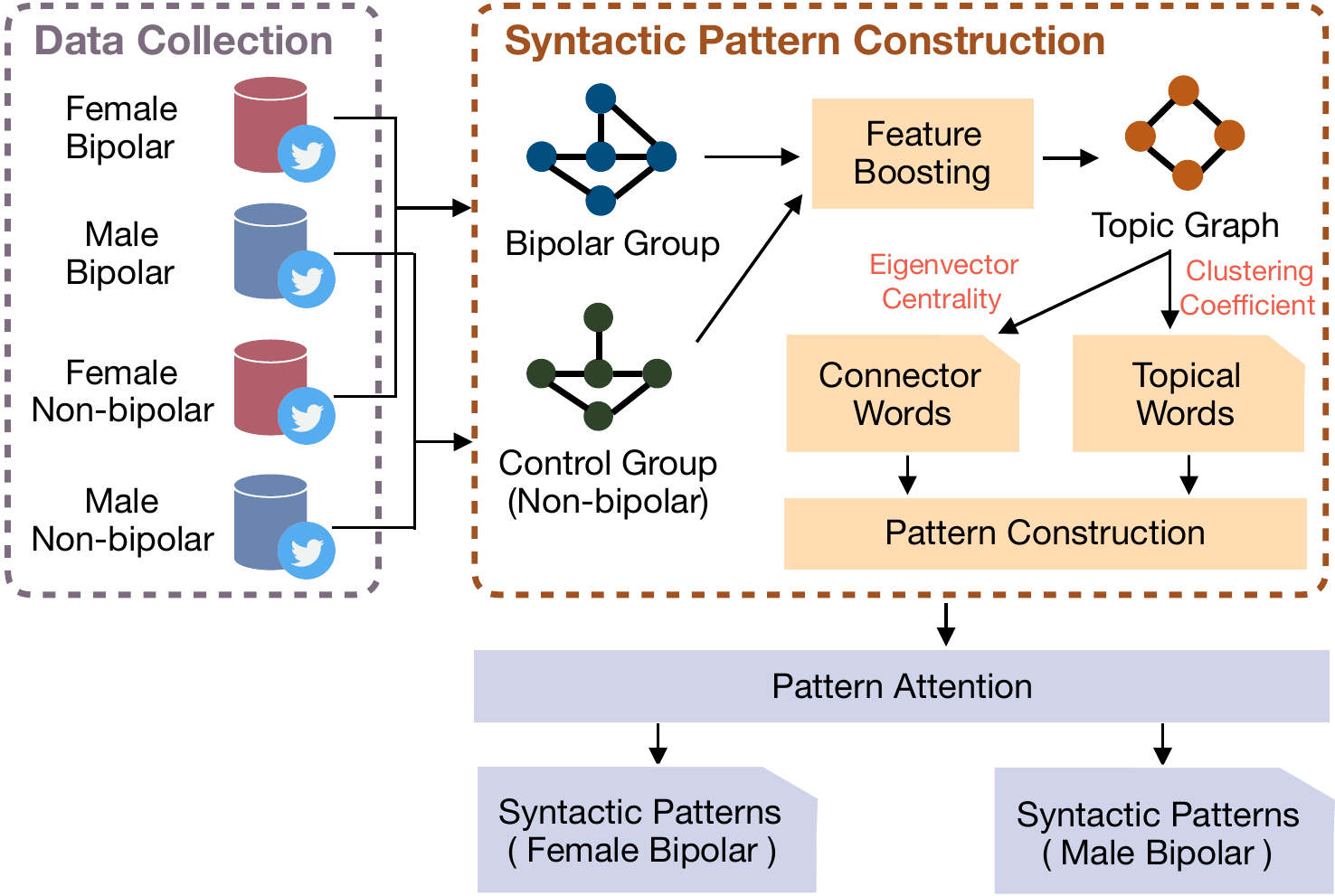}
    \caption{Framework}
    \label{fig:framework}
\end{figure}
\section{Syntactic Pattern Construction}

To dynamically learn syntactic patterns of word usage from BD, this work adapts the graph-based extraction algorithm in the emotion detection work of~\citet{saravia2018carer}.
By constructing a word relation graph, the hidden word relations are preserved to enrich the patterns in comparison to traditional lexicon-based approaches.
To enable gender characteristics, 
patterns are built separately for both gender, and integrated by the pattern attention mechanism.

\noindent
\textbf{Overview:} Given a group of BD users $\bm{U}$ and a control group $\hat{\bm{U}}$, the corresponding set of tweets within a time interval $\alpha$ for each group are denoted as $\Gamma^{(\alpha, \bm{U})}$ and $\Gamma^{(\alpha, \hat{\bm{U}})}$, respectively. 
A topic word graph $\tilde{G}^{(\bm{U})}$ is constructed by mapping two sets of tweets to a graph function, and syntactic patterns are derived through graph $\tilde{G}^{(\bm{U})}$. The objective is to employ these patterns as features to train a model to recognize whether a user is either likely to have BD or belongs to the control group.
The framework is shown in Figure~\ref{fig:framework}.

\noindent
\textbf{Graph Construction:} Before the pattern extraction phase, the word graph construction phases is defined below. 

\begin{definition}(Word Graph)
\label{def:word_graph}
A word graph $G$ is denoted as 
\begin{equation}
G = (\bm{V},\bm{E}, \bm{W})
\label{equ:word_graph}
\end{equation}
where $\bm{V}$ stands for a set of nodes representing words, $\bm{E}$ is the set of edges constituted by two nodes, and $\bm{W}$ represents the edge weights.
\end{definition}

A word graph $G$ is constructed using a set of sentences described as a sequence of tweets $\Gamma$ in our paper. Nodes in $G$ are linked together if two words are adjacent to each other. For example, if there exists word sequences such as ``I~am~...'', the word ``I'' and ``am'' are linked together. The edge weights are initialized as
$
 \forall w_{v_{i}, v_{j}} \in \bm{W}, \quad w_{v_{i}, v_{j}} = TF_{v_{i}, v_{j}}*DF_{v_{i}, v_{j}} 
$
where $TF$ is the frequency for each bi-gram and $DF$ is the probability of the co-occurrence among users indicating how many people use this bi-gram. We then build two word graphs (BD and control group) as $G^{(\bm{U})} = (\bm{V}^{(\bm{U})}, \bm{E}^{(\bm{U})}, \bm{W}^{(\bm{U})})$ and $G^{(\bm{\hat{U}})} = (\bm{V}^{(\bm{\hat{U}})}, \bm{E}^{(\bm{\hat{U}})}, \bm{W}^{(\bm{\hat{U}})})$, respectively.

As both of the graphs are built from online human-generated text medium, there are many common behaviors of language usage shared between the BD and control group. Such similar language usages may dominate the observation from graph analysis. We thus propose the \textbf{Feature Boosting} mechanism to highlight the target (BD) specific language usages.

\noindent
\textbf{Feature Boosting:} 
In the area of behavioral neuroscience, \citet{denson2018neural} studied the brain influence by alcohol with functional magnetic resonance imaging (fMRI) by measuring the difference between the correlation maps of the alcohol group and the placebo group.
Inspired by them, the graph-based feature boosting is designed 
to leverage the dis-similarity between BD and control graph and the topic graph is derived as follow: 

\begin{definition}(Topic Graph)
\label{def:topic_graph}
A topic graph for BD group $\tilde{G}^{(\bm{U})} = (\bm{V}^{(\bm{U})}, \bm{E}^{(\bm{U})}, \tilde{\bm{W}}^{(\bm{U})})$ is designed as:
\begin{equation}
\tilde{\bm{W}}^{({\bm{U}})} := \{max(0, w^{(\bm{U})}_{v_{i}, v_{j}} - w^{(\bm{\hat{U}})}_{v_{i}, v_{j}}) \}
, \quad
\forall w^{(\bm{U})}_{v_{i}, v_{j}} \in \bm{W}^{({\bm{U}})}
\label{equ:topic_graph}
\end{equation}
\end{definition}
For each word pair, the connection will be kept in the topic graph if the weight gap is greater than $0$. Rather than directly extracting features from two separate graphs, the concept of feature boosting can emphasize and preserve more representative information which is highly correlated to the target (BD) group.

\noindent
\textbf{Pattern Construction:} We now apply two types of graph analyses on the topic graph, namely \textit{eigenvector centrality} and \textit{clustering coefficient}, where each analysis helps to generate a set of connector words ($\bm{CW}$) and topical words ($\bm{TW}$) respectively. The thresholds $th_{ec}$ and $th_{cc}$ are empirically defined for centrality and coefficient scores. Nodes in the topic graph that satisfy the thresholds are kept. Given the two set of words, $\bm{CW}$ and $\bm{TW}$, permutations containing these two types of words, by the rule of two connector words with one topical word, are served as templates to be used to exhaustively search for syntactic patterns from the training dataset. Finally, the $\bm{TW}$ component of each pattern is replaced with the wildcard~*, which increases pattern coverage when extracting features on testing time and also enables the content independent capability.

\section{Pattern Attention}
Since not all patterns are equally important, we propose the pattern attention mechanism, which is constructed by intra-group and inter-group attention, to arrange the importance of the patterns. 

\noindent
\textbf{Intra-group attention}, which compares within-group difference between each pattern, is calculated using three measures.
\begin{enumerate*}
    \item \textbf{Pattern Frequency (PF)} refers to the probability of a pattern occurring in the target group. The higher the frequency, the more essential is the pattern in that group.
    \item \textbf{User Frequency (UF)} refers to the probability of the co-occurrence of a pattern among users in the group and ensures that a pattern is used by the majority of users in that group.
    \item \textbf{Diversity (DIV)} refers to the number of unique topical words that a pattern can match. To ensure coverage of and tolerance toward unseen words, the useful patterns must be able to capture a wider range of topical words in a different collection.
\end{enumerate*}

\noindent
\textbf{Inter-group attention} is another essential indicator to adjust the significance of a pattern by examining the condition in which it is used in different groups. Considering the gender-based dimension is associated with the behaviors of self-disclosure and the linguistic expressions, we design \textbf{Contrast Gender Frequency (CGF)} to quantify the differences between pattern usage of females and males. The \textbf{CGF} is the normalized value of the \textbf{PF} across gender subgroups. Not only do we evaluate the gender-based effects, but we take the mental health status into consideration.
An additional measure, \textbf{Contrast Onset Frequency (COF)}, derived from the same concept of \textbf{CGF} by replacing the gender subgroups with bipolar and control groups, is to emphasize the mental illness concerns on pattern attention. The inter-group attention mechanism can reward the pattern in the group of which it is more representative.

The pattern attention score is the product of the above five assessment methods, and the proposed syntactic patterns are sorted with their scores for the target group.

\section{Recognition Performance}

\begin{figure}
\centering
\includegraphics[width=\linewidth]{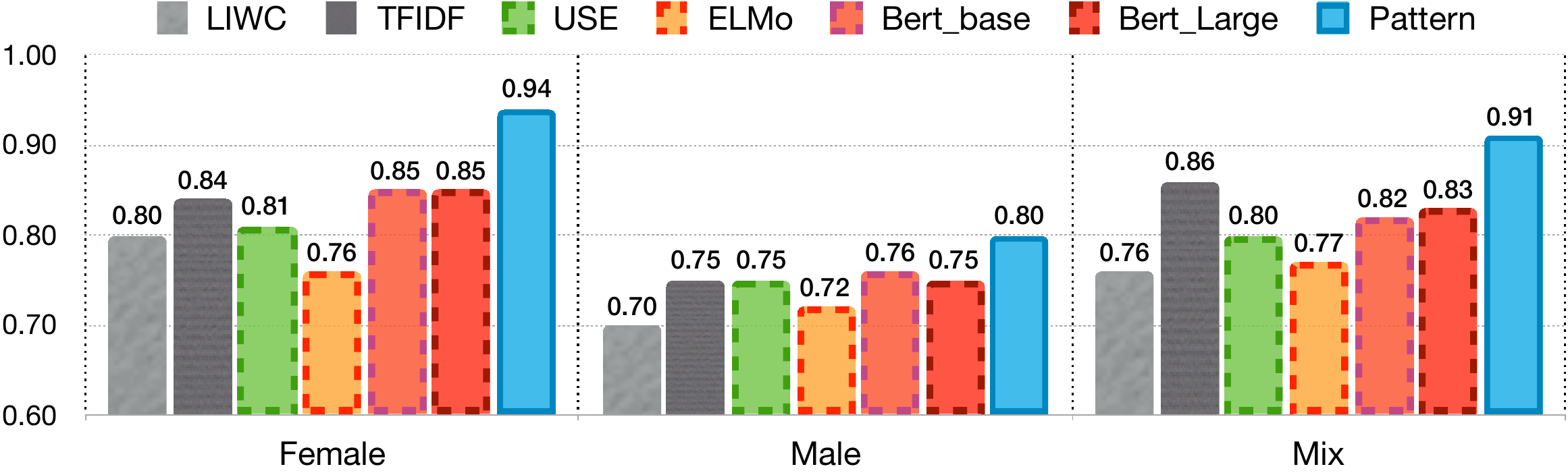}
\caption{Overall Performance (F1)}
\label{fig:Performance_bar}
\end{figure}

\begin{table}[]
    \caption{Precision for BD Recognition}
    \label{tab:precision_bd}
    \footnotesize
    \centering
    \begin{tabular}{c|c|c|c|c|c|c|c}
    \hline
     & \textbf{LIWC} & \textbf{TF-IDF} & \textbf{USE} & \textbf{ELMo} & \textbf{BERT-b} & \textbf{BERT-l} & \textbf{Pattern}  \\
    \hline
        Female & 0.76 & \textbf{0.85} & 0.70 & 0.75 & 0.81 & 0.75 & \textbf{ 0.95} \\
        Male & 0.71 & 0.76 & 0.70 & \textbf{0.80} & 0.79 & 0.62 & \textbf{0.88} \\
        Mix & 0.76 & \textbf{0.87} & 0.76 & 0.78 & 0.80 & 0.75 & \textbf{0.96} \\
    \hline
    \end{tabular}
\end{table}

\subsubsection*{\textbf{Experimental Setup}}
The proposed syntatic patterns were performed as a normalized bag of pattern features and first compared with with two traditional baseline method: 
\begin{enumerate*}
    \item \textbf{TF-IDF featured model}~\cite{coppersmith2014quantifying,chang2016subconscious}: uses $1$ and $2$ grams features, where a document represents a user; and
    \item \textbf{LIWC featured model}~\cite{coppersmith2014quantifying, sekulic2018not}: $64$ categories were selected from the LIWC affect lexicon, including syntactic features, topical features (e.g., work and friends), and psychological features (e.g., emotions and social context).
\end{enumerate*}
A random forest algorithm was applied to leverage the features with a tree size of $128$ as well as a minimum number of samples required to be at a leaf node of $3$. In addition, the following state-of-the-art pre-training laguage models were also compared to our method:  \textbf{ELMo}~\citep{Peters:2018},  \textbf{BERT}~\citep{devlin2018BERT}, and Universal Sentence Encoder (\textbf{USE})~\citep{cer2018universal} from \textit{Google}, as these models need fewer data to be fine-tuned. The context-aware word representations were retrieved using the pre-trained model on \textit{TFhub}~\footnote{\url{https://tfhub.dev/}} and summarized to a fixed-length sentence encoding vector with the element-wise mean of the representations at each word position.
The vector representations were fine-tuned using a dense neural network with three hidden layers ($512$, $128$, $64$) with a dropout rate of $0.3$ for each layer and the \textit{Adam} optimizer.

The proposed syntactic patterns were evaluated with F1 score, leveraging both precision and recall, as the mis-classification may result in serious consequence in clinical practice.

\noindent
\textbf{Gender-separated Case}:
Motivated by the divide and conquer strategy, as the gender differences have a great influence on mental disorder, we first examined the proposed syntactic patterns in a gender-separated case, which is also similar to real life clinical situation.
Given a set of BD and control group users based on gender, the recognition performance of the gender-specific \textbf{Pattern} outperformed other baseline methods for both gender groups (shown in Figure~\ref{fig:Performance_bar}).
We also observed that the performance of females was higher than males in every method, due to the lack of male users.
It indicates that if the gender information is given, which is common in clinical case, it is possible to acquire a higher performance by carefully select the (gender specific) features.

\noindent
\textbf{Gender-mixed Case}:
In this experiment, female and male BD datasets were merged as a united BD dataset (N=349) as well as the control dataset.
The gender-specific patterns were combined as gender-enriched patterns which outperformed the baselines, shown in Figure~\ref{fig:Performance_bar} (right). This indicated that patterns were able to learn both female and male features by integrating gender-separated patterns, the performance could be benefited even when there were no available gender information for testing dataset.

\noindent
\textbf{Assistive Ability}:
Recall the aim of our study, which is to build assistive syntactic patterns for the healthcare practitioners. The precision for BD recognition were highlighted in Table~\ref{tab:precision_bd}, where the proposed \textbf{Pattern} outperformed all baseline models. As an assistive tool for the BD assessments, the proposed syntactic patterns could be served as an reliable reference in clinical practice.

For the above experiments, our proposed syntactic patterns outperform several baselines, including traditional lexicion-based methods and the state-of-the-art models.  
The \textbf{LIWC} method relies on a predefined dictionary, which may face an out-of-vocabulary problem, especially for informal texts. For \textbf{TF-IDF}, we observed that it outperformed several baselines, since it learned the content related n-grams from datasets. While the patterns rely less on keywords (shown in Figure~\ref{fig:top_match_pattern}), which is more content independent. 
Most of existing pre-training language models, such as \textbf{USE}, \textbf{ELMo} and \textbf{BERT}, provide the sentence-level contextual representation, which may suffer from the document tasks, where a user could post many sentences during the period.
Although these pre-training language models have outstanding performances, the interpretability may arise as an issue of concern, especially for clinical analysis.

\section {Analysis and Discussion}

\subsubsection*{\textbf{Syntactic Patterns}}
\begin{figure} 
    \centering
    \includegraphics[width=1\linewidth]{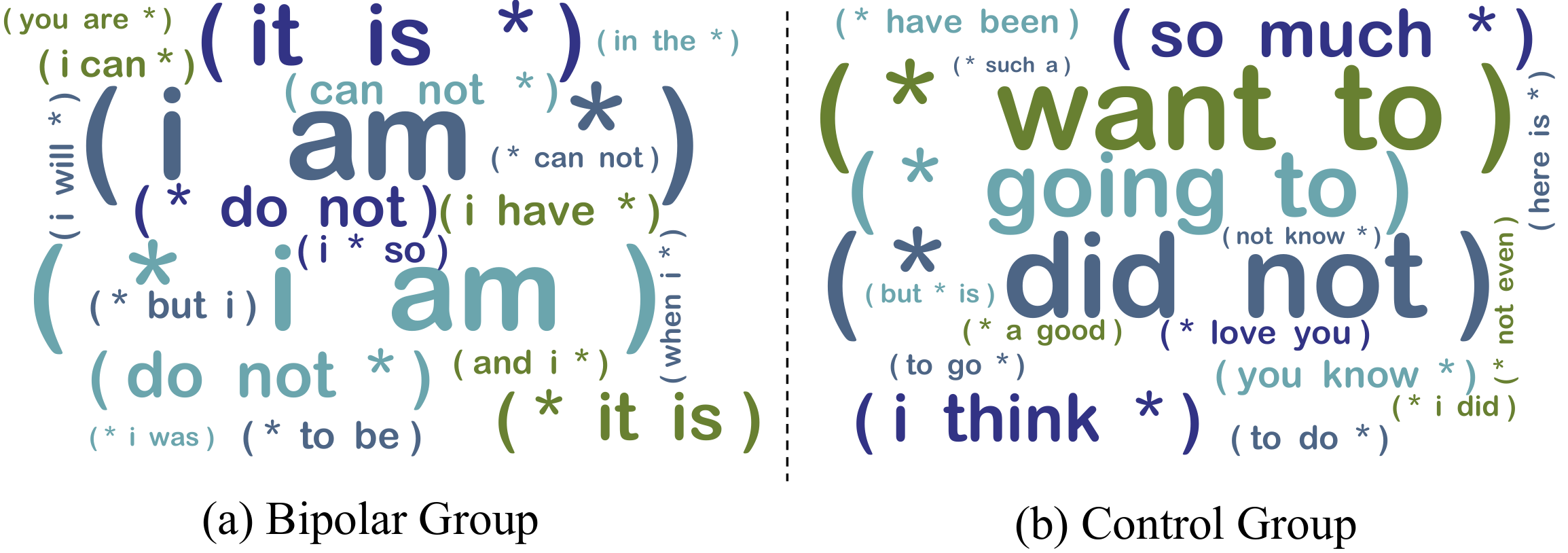}
    \caption{Top patterns}
    \label{fig:top_match_pattern}
\end{figure}

To observe what the proposed method captured, the most frequent patterns were extracted for the BD and the control groups in Figure~\ref{fig:top_match_pattern}. A previous study on psycholinguistics~\citep{ireland2014natural} has shown that the increased use of first-person singular pronouns is a common symptom of people suffering from some form of mental illness. 
Whereas this behavior was $1.7$-times more common in our dataset. 
By leveraging the expressiveness of the patterns, it is possible to observe that the patterns most frequently used by the BD group notably contain the first-person singular pronoun, ``i'', and topical words that express negative emotions, such as ``trying'', ``tired'', and ``never''. 
Pyszczynski et al.~\cite{pyszczynski1987self} reported that depressed people tend to focus on themselves rather than engaging with others, which could explain the difference in the word usage between the groups. 
Al-Mosaiwi et al.~\cite{al2018absolute}  reported that absolute words, such as ``always'' and ``never'', are also reliable markers for diagnosing mental illness. 

Overall, in this study, the word usage of the patterns for the control group showed more positive topical words (e.g., fun, please) and less use of first-person pronouns and absolute words, while the pattern for the BD group tended to show less social words (e.g., we, \textit{@mention}), more transition words (e.g., but and cause), and regret.

\subsubsection*{\textbf{Gender Differences in Patterns}}
\begin{figure} 
    \centering
    \includegraphics[width=1\linewidth]{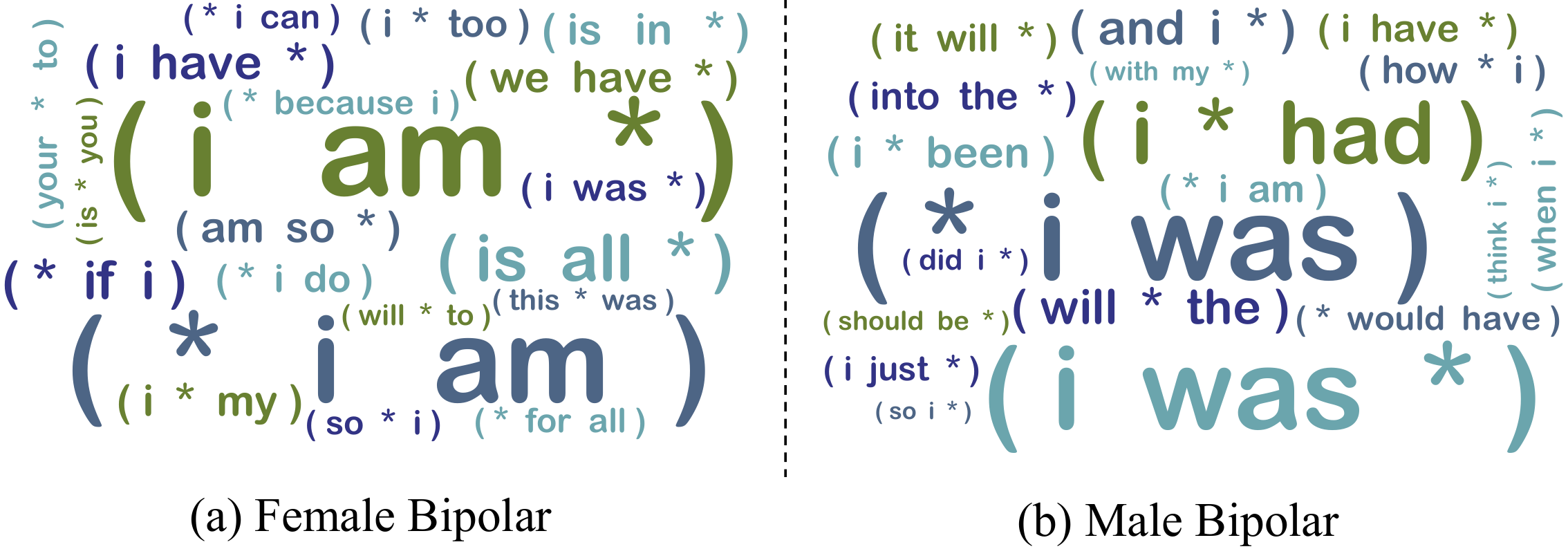}
    \caption{Gender top patterns}
    \label{fig:gender_top_match_pattern}
\end{figure}

The proposed syntactic patterns were enriched with gender-based features by integrating two separated gender-specific patterns. To better understand the gender differences in word usage, we separately trained the recognition models for gender with only the corresponding gender-specific features alone and studied the most influential patterns from the models.
As shown in Figure~\ref{fig:gender_top_match_pattern}, consistent with the previous section, there was a similar observation for the first-person pronouns, which indicates that people with BD were more focus on themselves. 
Furthermore, we observed that female BD users tended to express with the present tense (e.g.,  ``i~am~*'', ``i~have~*'') while male BD users used the past tense (e.g., ``i~was~*'', ``i~had~*'') more often. 
The statistical t-test was following conducted for the BD group to verify the significance level among different tenses~\footnote{The words represented for each tense were retrieved from the LIWC~\cite{pennebaker2001linguistic} dictionary.}, and it experimentally showed that there were significant differences for the future tense ($p{=}0.037$), present tense ($p{<}0.001$) and past tense ($p{=}0.013$). 
\citet{munmun2017gender} also reported the similar observations, and we additionally demonstrated the difference of the past tense for gender. Overall, the gender differences of word usage could be observed from our approach and, more importantly, the word usage could be demonstrated in an n-gram or skip-gram pattern manner which is more context interpretable.

\section{Limitation and Future Work}
There are some limitations in this study.
First, we employed a data collection method where users who self-reported as being diagnosed with BD are identified. This method is difficult to verify the authenticity of the self-disclosure statement. However, as previous studies~\cite{coppersmith2014quantifying, munmun2017gender} mentioned, it is unlikely that users would self-disclose that they are diagnosed with mental illness they do not have. 
We expect to retrieve different samples of people who are clinically diagnosed as having mental illness to conduct further assessments.

We also admit that manually labeling gender for BD group is a highly time-consuming process which makes the reproducibility more laborious. But this can assure the quality of information and extend our ability to infer BD users' gender for the future tasks.
\bibliographystyle{ACM-Reference-Format}
\bibliography{main}

%

\end{document}